\documentclass[12pt]{article}
\usepackage{cite,amsmath,amssymb}
\usepackage[utf8]{inputenc}
\usepackage[right=1.5cm,left=1.5cm,top=2.5cm,bottom=4.9cm,a4paper,twoside]
{geometry}
\usepackage{amsxtra}
\usepackage{amsfonts}

\newcommand{\be}{\begin{equation}}
\newcommand{\ee}{\end{equation}}
\newcommand{\bea}{\begin{equation}\begin{aligned}}
\newcommand{\eea}{\end{aligned}\end{equation}}
\renewcommand{\phi}{\varphi}

\newcommand{\corrbase}[1]{\ensuremath{\langle \, #1 \, \rangle }}

\newcommand{\corrk}[3]{ \ensuremath{\corrbase{ #1_{ \vec{#2}} #1_{\vec{#3}} }} }

\newcommand{\dphi}{\ensuremath{\delta\phi}}

\newcommand{\chik}[1]{ \ensuremath{\chi_{\vec{#1}}} }

\newcommand{\kmes}[1]{\ensuremath{\frac{d^3 #1}{(2\pi)^3} \, }}

\newcommand{\random}[1]{ \ensuremath{ \hat{a}_{\vec{#1}} } }

\newcommand{\xv}{\ensuremath{\vec{x}}}
\newcommand{\yv}{\ensuremath{\vec{y}}}

\newcommand{\None}{\ensuremath{N_\phi}}
\newcommand{\Ntwo}{\ensuremath{N_{\phi\phi}}}
\newcommand{\Nthree}{\ensuremath{N_{\phi\phi\phi}}}

\begin{document}

\thispagestyle{empty}
\vspace*{.2cm}
\noindent
HD-THEP-10-10  \hfill 18 May 2010
\\
BI-TP 2010/15

\vspace*{1.0cm}

\begin{center}
{\Large\bf Inflationary Infrared Divergences:\\[.2cm] 
Geometry of the Reheating Surface vs. $\delta N$ Formalism}
\\[1.5cm]

{\large C.T. Byrnes$^{\,a}$, M.~Gerstenlauer$^{\,b}$, 
A. Hebecker$^{\,b}$, S. Nurmi$^{\,b}$, G. Tasinato$^{\,b}$}\\[6mm]
{\it
$^{a}$~Fakult\"at f\"ur Physik, Universit\"at Bielefeld, Postfach 100131, 
33501 Bielefeld, Germany\\[3mm]
$^{b}$ Institut f\"ur Theoretische Physik, Universit\"at Heidelberg, 
Philosophenweg 16 und 19, D-69120 Heidelberg, Germany}
\\[.5cm]
{\small\tt (byrnes@physik.uni-bielefeld.de {\it and} m.gerstenlauer,\,
a.hebecker,\,s.nurmi,\,g.tasinato@thphys.uni-heidelberg.de)}
\\[2.0cm]

{\bf Abstract}
\end{center} 
We describe a simple way of incorporating fluctuations of the Hubble 
scale during the horizon exit of scalar perturbations into the $\delta N$ 
formalism. The dominant effect comes from the dependence of the 
Hubble scale on low-frequency modes of the inflaton. This modifies the 
coefficient of the log-enhanced term appearing in the curvature spectrum at 
second order in field fluctuations. With this modification, the relevant 
coefficient turns out to be proportional to the second derivative of the 
tree-level spectrum 
with respect to the 
inflaton $\phi$ at horizon exit. A logarithm with precisely the same
coefficient appears in a calculation of the log-enhancement of the curvature 
spectrum based purely on the geometry of the reheating surface. We take this
agreement as strong support for the proposed implementation of the $\delta N$
formalism. Moreover, our analysis makes it apparent that the log-enhancement 
of the inflationary power-spectrum is indeed physical if this quantity is
defined using a global coordinate system on the reheating surface (or any
other post-inflationary surface of constant energy density). However, it can  
be avoided by defining the spectrum using invariant distances on this 
surface.

\newpage

\section{Introduction}
It is well known that curvature perturbations created during cosmological 
inflation \cite{infl,Riotto:2002yw} are affected by infrared (IR) divergences 
\cite{Lyth:2007jh,ird} (see \cite{recent_ird} for some recent 
discussions). These divergences are closely related to the familiar divergence 
of the scalar-field correlator in de Sitter space~\cite{dS}. While, at leading 
order, the divergence can be absorbed into the definition of the background 
and is hence unobservable, higher orders in the curvature perturbation lead 
to corresponding log-enhanced corrections to the power spectrum. Since the IR 
cutoff appearing in these logarithms is provided by the size of the observed 
universe (rather than by, for example, the size of the universe created by 
inflation), these logarithms are, however, not particularly large in 
practice~\cite{Lyth:2007jh}. Nevertheless, it is conceivable that the 
power spectrum is measured by a very late observer, who has access to the 
entire region of the universe created in `our' inflationary patch.\footnote{
Of 
course, this possibility is in practice limited by the presently observed 
dark energy or cosmological constant.
}
For such a `late' observer, the infrared logarithms can be extremely large 
and it is an interesting question of principle how to achieve consistency 
between his and our (i.e. the `early' observer's) measurement of the power 
spectrum.

We approach these issues starting from the $\delta N$ formalism \cite{deltaN,
lr}.\footnote{
The 
physical importance of higher-order terms in the $\delta N$ formalism 
was first appreciated in \cite{lr}.
} 
It turns out that the problem can be resolved rather easily if a simple 
modification of the $\delta N$ formalism, which takes fluctuations of the 
Hubble scale during slow-roll inflation into account, is implemented. In 
essence, this modification consists of treating the Hubble scale, which 
defines the normalization of the scalar-field correlator, as a function of the 
perturbed background value of $\phi$ relevant at the time of horizon exit of a 
given mode. Since this background perturbation depends only on modes with a 
smaller wave number, our proposal can be implemented in an unambiguous and 
straightforward way. 

The Hubble-scale fluctuations discussed above lead to slow-roll suppressed 
but log-enhanced contributions to the power-spectrum, similar to the 
familiar `c-loop' effects of the $\delta N$ formalism. Combining both 
these
contributions, the coefficient of the first log-enhanced correction to the 
power spectrum takes a very simple form. It is essentially given by the 
second derivative in $\phi$ of the leading order power spectrum, 
$N_\phi(\phi)^2H(\phi)^2/(2\pi)^2$. Here the argument $\phi$ is the
value of the inflaton corresponding to the horizon exit of the 
wave number $k$ under consideration.

The physical significance of this proposal can be understood as follows:
Consider the reheating surface (or any other surface of constant energy 
density after the end of inflation). Disregarding vector and tensor modes, 
the geometry of this surface, parameterized by coordinates 
$\vec{y}=(y^1,y^2,y^3)$, can be characterized by a single scalar
function $\zeta(\vec{y})$, known as the curvature perturbation. The
power spectrum may be defined as the logarithmic derivative of the 
correlation function $\langle\zeta(\vec{x})\zeta(\vec{x}+\vec{y})\rangle$ 
with respect to $y=|y|$. Alternatively, a closely related spectrum based 
on a fixed invariant distance $s$ between pairs of points in the correlator
can be defined. This latter power spectrum is an entirely local 
quantity and its expectation value does not depend on the size of the 
region over which the correlator is measured (i.e. on the `age' of the 
observer). The two spectra, which we denote by ${\cal P}_\zeta(y)$ and 
$\tilde{\cal P}_\zeta(s)$, are related by ${\cal P}_\zeta(y)=\langle{\tilde{\cal 
P}}_\zeta(ye^{\bar{\zeta}})\rangle$, where $\bar{\zeta}$ is a coarse-grained 
value of $\zeta$ relevant for the distance-measurement between a given 
pair of points. In the relation 
between the two spectra, the averaging is over the potentially large 
observed region, with large expectation values of $\bar{\zeta}$ and 
$\bar{\zeta}^2$ in the case of a late observer. This leads to a 
log-enhancement of ${\cal P}_\zeta(y)$ which is easily seen to be precisely 
the log-enhancement found earlier in (our implementation of) 
the $\delta N$ formalism. 

On the one hand, this agreement provides support for the implementation of 
the $\delta N$ formalism that we advocate. On the other hand, it makes the
`physical reality' of large logs from IR divergences particularly clear:
The log-enhancement arises due to the use of global coordinates in a very
large region, where these deviate significantly from the invariant distance.
It can be avoided if one measures the power spectrum $\tilde{\cal P}_\zeta(s)$,
which is defined using a two-point correlation function based on the 
invariant distance between each pair of points appearing in the spatial 
average.

While our work was being finalized, Ref. \cite{Giddings:2010nc} appeared, 
which overlaps with part of our analysis. We will comment on this in more
detail at the end of Sec.~\ref{sec:reheating}.

\section{Hubble scale fluctuations in the $\delta N$ formalism}
During inflation, the amplitude of scalar field fluctuations is
controlled by the Hubble parameter $H$. The latter is usually evaluated at 
the value $\phi(t_k)$ of the classical homogenous solution $\phi$,
where $t_k$ is the time of horizon exit of the mode $k$: $H=H(\phi
(t_k))$. However, this approach does not account for the fact that 
perturbations with wavelength larger than $k^{-1}$ have already left
the horizon. These perturbations  modify the value of the scalar field 
relevant for the mode $k$ and need to be taken into account. 

We do so by writing the scalar field perturbation in a quasi-de-Sitter 
background as
\bea
 \label{eq:ansatz_real_space}
 \dphi (\xv) & = \int \kmes{k} \frac{e^{-i\vec{k}\vec{x}}}{\sqrt{2k^3}} 
  \; \;H \left( \phi (t_k) + \delta \bar{\phi}(\vec x) \right) \; \;
  \random{k}\,, 
\eea 
where
\bea
 \delta \bar \phi (\xv)  & = \int\limits_{l \ll k} \kmes{l} 
 \frac{e^{-i\vec{l}\vec{x}}}{\sqrt{2l^3}}\; H ( \phi (t_l) ) \; \; \random{l}\,,
\eea
where $\random{k}$ is a normalized Gaussian random variable~\footnote{It
satisfies the relation $\langle
\hat{a}_{\vec k} \hat{a}_{\vec p}
 \rangle
\,=\,
(2\pi)^3 \delta^{(3)}(\vec{k}+\vec{p})\,.$}. In writing
eq. (\ref{eq:ansatz_real_space}) and in the rest of the paper, we
only include the leading order contributions to the fluctuations and
neglect slow-roll corrections that are not enhanced by potentially
large logarithms. 
The $\delta \bar \phi(\vec x)$ contribution in the  argument of
the Hubble parameter accounts for the backreaction of long wavelength modes
on the scalar fluctuations $\delta \phi(\vec x)$ generated inside the horizon. 

In order to analyze their effect, we expand the Hubble parameter
up  to second order in $\delta \bar \phi$, 
\be
 \label{eq:scal_field_fluct}  
 \dphi (x) = \int \kmes{k} e^{-i \vec{k}\vec{x}} \chik{k} \left[ 1 +
   A_1 \int\limits_{l \ll k} \kmes{l} \, e^{-i \vec{l}\vec{x}}
   \chik{l} \, + \;  A_2 \int\limits_{l, m \ll k} \kmes{l} \kmes{m}
   e^{-i (\vec{l} + \vec{m}) \vec{x}} \chik{l} \, \chik{m} \right] 
\ee
where we have introduced the Gaussian field 
\be
\chik{k} = \frac{H(\phi (t_k))}{\sqrt{2k^3}} \; \random{k}\,.
\ee
We use the abbreviations $A_1 = H_\phi/H$ and $A_2 = 
H_{\phi\phi}/(2H)$ for coefficients involving the first and second
derivative of $H$ with respect to $\phi$. Note that as a result of the 
corrections proportional to $A_1$ and $A_2$, the scalar fluctuation
in Fourier space, $\delta \phi_{\vec k}$, are not Gaussian
at Hubble exit. A similar idea was considered to second order in the scalar 
field perturbations, in the context of the tree-level bispectrum in 
\cite{Allen:2005ye}. Note that we neglect the quantum mechanically
generated non-Gaussianity of the fields which is present at horizon 
exit \cite{Maldacena}.

We now turn to the curvature fluctuation $\zeta$ evaluated on a
surface of uniform energy density (for our purposes, the reheating surface).
The $\delta N$ formalism calculates $\zeta$ starting from scalar fluctuations
on a flat surface,
\be\label{dnexp}
\zeta(\vec{x})\,=\,N(\phi+\delta\phi(\xv))-N(\phi)\,=\,N_{\phi} \delta 
\phi(\vec{x})\,+\,\frac12\, N_{\phi\phi}\delta
\phi(\vec{x})^2\,+\,\frac{1}
{6}N_{\phi\phi\phi}\delta\phi(\vec{x})^3\,+\cdots\,.
\ee
Here $N_\phi$, $N_{\phi\phi}$ are derivatives of the number of e-folds $N$
as a function of $\phi$, evaluated on the classical background trajectory.
The resulting two point function for curvature perturbations in Fourier
space is given by
\be
\label{eq:delta_N}
\corrk{\zeta}{k}{p} = \None^2 \corrk{\dphi}{k}{p}  
+ \None \Ntwo \corrbase{ \dphi_{\vec{k}}\!\left(\dphi^2\right)_{\vec{p}} }
+ \frac{\Ntwo^2}{4} \corrk{\left(\dphi^2\right)}{k}{p} 
+ \frac{\None\Nthree}{3} \corrbase{ \dphi_{\vec{k}}
\!\left(\dphi^3\right)_{\vec{p}} }.
\ee
Note that correlators of odd powers of $\dphi$ are in general non-zero since
$\dphi_{\vec k}$  is not Gaussian. However, one may check that
nevertheless $\corrbase{\dphi_{\vec{k}} } = 0 $. 
 We have not set $<\zeta>=0$, though  making this
 choice  would not change in an essential way our results (we will discuss this in more detail later).

We now define the spectrum ${\cal P}_{\chi}$ through $\langle\chi_{\vec{k}} 
\chi_{\vec{p}}\rangle\,=\,(2\pi)^3\,\delta^3(\vec{k} +\vec{p}\,)\,2 \pi^2 
{\cal P}_{\chi} (k) /k^3$. This implies that ${\cal P}_{\chi}(k)=(H/2\pi)^2$.
The spectrum of curvature perturbations ${\cal P}_\zeta$ is defined 
in an
analogous way.
  It follows from a straightforward evaluation of 
eq.~(\ref{eq:delta_N}) (see Appendix A for details) that
\bea
  \label{eq:deltaN_final}
  \mathcal{P}_{\zeta}(k) & =  \mathcal{P}_\chi \left\lbrace \None^2 +
    \mathcal{P}_\chi \ln(kL) \left[ \left( A_1^2 + 2 A_2 \right)
      \None^2 + 4 A_1 \None\Ntwo + \Ntwo^2 + \None\Nthree \right]
  \right\rbrace \\ 
   & = \mathcal{P}_\chi \None^2 + \mathcal{P}_\chi \ln(kL) \;
   \frac{1}{2}  \frac{d^2}{d\phi^2}\left( \None^2 \mathcal{P}_\chi
   \right)\,. 
\eea
Here it is crucial that the IR cutoff $L$ is set by the size of
the region in which the correlator is measured. As mentioned before,
we take it for granted that perturbations on even larger scales can be 
absorbed in the background and are hence unobservable. In other words, 
the classical evolution starts with the horizon exit of modes of scale
$k\sim 1/L$. Our main interest is the dependence of the spectrum on
the size $L$ of the observed region.

We can express the second derivative along the scalar field, appearing
in equation (\ref{eq:deltaN_final}), in terms of  slow-roll parameters. 
We define as usual $\epsilon\,=\,V_\phi^2/\left(2 V^2\right)$,
 $\eta\,=\,V_{\phi\phi}/V$,
$\xi^2\,=\,V_\phi V_{\phi\phi\phi}/V^2$, and we set $M_{P}^2 = 8 \pi$.
 Using the well known relations $n_\zeta-1=2
\eta- 6 \epsilon$
for
the spectral index, and
$\alpha_\zeta\,=\,16 \epsilon \eta-24 \epsilon^2-2 \xi^2$ for its running,
i.e. $\alpha_\zeta\,\equiv\,d n_\zeta / d \ln{k}$ (all
quantities evaluated at Hubble exit) we find
\be\label{srpars}
\frac{d^2}{d \phi^2}\left( N_{\phi}^2 {\cal P}_\chi\right)
\,=\,\frac{{\cal P}_{\chi}\,N_{\varphi}^4}{4}\,\left[
 \left(\eta-2\epsilon\right)\,\left(n_\zeta-1\right)
\,+
\,\left(n_\zeta-1\right)^2\,+\,\alpha_\zeta
\right]
\ee
This quantity vanishes in the case of scale invariance, i.e.~when
$n_\zeta=1$ and $\alpha_\zeta=0$. In this case logarithmic corrections
to the power spectrum drop out, at leading order in  the slow-roll
expansion. We will discuss an interpretation of this result in the
next section.

At our level of accuracy, we can rewrite eq.~(\ref{eq:deltaN_final}) as
\be
\mathcal{P}_{\zeta}(k)=\mathcal{P}_\chi \None^2 + \langle\phi^2\rangle_{1/k}\;
\frac{1}{2}  \frac{d^2}{d\phi^2}\left( \None^2 \mathcal{P}_\chi\right)\,,
\ee
where
\be 
\langle\delta\phi^2\rangle_{1/k}=\int\limits_{L^{-1}}^k \kmes{k'}
\frac{H^2}{2k'^3} = \mathcal{P}_\chi \ln(kL)
\label{df2}
\ee
is the expectation value of $\delta\phi^2$ measured on a length scale
$1/k$. We see that the log-enhanced correction takes the suggestive form
of the second term in a Taylor expansion. The physical meaning of this 
structure will be clarified below. 

We note that the integration in eq. (\ref{df2}) was performed
assuming that $\phi$ has an exactly scale invariant spectrum. More generally, if the power spectrum 
of $\phi$ has a constant spectral index $n_\phi$, the $\ln(kL)$ term
above should be replaced by
\be
\ln(kL)\rightarrow
\frac{1}{n_{\phi}-1}\left(1-\left(kL\right)^{-(n_{\phi}-1)}\right). 
\ee
However, provided that we do not consider an exponentially large range 
of scales and the field fluctuations are reasonably close to scale 
invariant, then $|n_\phi-1|\ln(kL)\ll1$, and the more general result
above is well approximated by $\ln(kL)$. Since the running of the
spectral index is expected to be of order $(n_\phi-1)^2$, further 
corrections associated with this running are suppressed 
by a similar argument.

\section{Infrared divergences from the geometry of the reheating surface}\label{sec:reheating}

We now provide a physical interpretation for the log-enhanced correction
to the power spectrum given in eq. (\ref{eq:deltaN_final}). An observer, 
in order to make a measurement, specifies a coordinate system on the 
reheating surface (or on any other surface of constant energy density after
the end of inflation). This can be done as follows. Choose a coordinate 
system where slices of constant time $t$ have uniform energy density.
 Neglecting vector and tensor modes, we write the metric as
\be
d s_4^2\,=\,-d t^2+a^2 (t) \,e^{2 \zeta({x})}\,\delta_{ij}\,d x^i d x^j
\,.
\label{metans}
\ee
We use conventions where reheating occurs at time $t_f$, at which
the homogeneous scale 
factor $a(t_f)=1$. This leads to the following metric on the 
three-dimensional reheating surface:
\be
d s_3^2\,=\,e^{2 \zeta(x)}\,\delta_{ij}\,d x^i d x^j\,.
\ee

Important consequences for the power spectrum of $\zeta$ can be
derived just from the geometry of the reheating surface specified
above. To see this, let us define the power spectrum as the
logarithmic derivative of the two point function,
\be
{\cal P}_\zeta(y)\,\equiv\,\frac{d}{d \ln{y}}\,\frac{1}{2}\,\langle 
\left(\zeta(\vec{x})-\zeta(\vec{x}+\vec{y})\right)^2 \rangle =-
\frac{d}{d \ln{y}}\,\langle \zeta(\vec{x}) \,\zeta(\vec{x}+\vec{y}) \rangle\,,
\label{psrs}
\ee
where $y$ is the length of a coordinate vector on the reheating
surface, $y^2=(y^1)^2+(y^2)^2+(y^3)^2$. It is not difficult to see 
that eq. (\ref{psrs}) gives the correct  value for the power spectrum
associated with the fluctuations of a slowly-rolling scalar in quasi-de
Sitter background (see Appendix B). 

Alternatively, we can define the power spectrum averaging over pairs
of points separated by a fixed invariant distance $s$, 
\be
\tilde{\cal P}_\zeta(s)\,\equiv\,\frac{d}{d \ln{s}}\,\frac{1}{2}\,\left\langle 
\left(\zeta(\vec{x})-\zeta(\vec{x}+\vec{y}(s)\right)^2
\right\rangle\,.
\ee
Here $\vec{y}(s)$ denotes a coordinate vector with invariant length
$s$. We have introduced a tilde to distinguish this spectrum from 
eq.~(\ref{psrs}). It is clear that long-wavelength background 
fluctuations of $\zeta$, which affect only the parameterization but
not the physics of any local patch of the reheating surface, are
irrelevant for this new spectrum. In other words, $\tilde{\cal P}_\zeta(s)$ 
is a purely local quantity which, by its very definition, cannot
depend on the size $L$ of the observed region.

Defining $\bar{\zeta}$ as the average value of $\zeta$ characteristic 
of a small region containing a particular pair of points $\vec{x}$ and 
$\vec{x}+\vec{y}(s)$, we can write
\be
\tilde{\cal P}_\zeta(s)\,\equiv\,\frac{d}{d \ln{s}}\,\frac{1}{2}\,\left\langle 
\left(\zeta(\vec{x})-\zeta(\vec{x}+\vec{e}se^{-\bar{\zeta}}\right)^2
\right\rangle\,.
\ee
Here $\vec{e}$ is a coordinate unit vector. We see that the two
spectra differ only by the selection of pairs of points in the averaging 
procedure. In one case, pairs with fixed coordinate distance $y$, in the
other case pairs with fixed invariant distance 
\be
s\,=\,e^{\bar{\zeta}}\,y
\ee
are chosen. Hence we can express ${\cal P}_\zeta$ in terms of the
locally defined spectrum $\tilde{\cal P}_\zeta$ through
\be\label{zell}
{\cal P}_\zeta (y)=-\frac{d}{d \ln{y}}\,\langle \zeta(\vec{x}) \,
\zeta(\vec{x}+\vec{y}) \rangle=
-\frac{d}{d \ln{s}}\,\langle 
\zeta(\vec{x})\,\zeta\left(\vec{x}+\vec{e}\big(ye^{\bar{\zeta}}\big)
e^{-\bar{\zeta}}\right)
\rangle=
\langle\tilde{\cal P}_\zeta (ye^{\bar{\zeta}})\rangle\,.
\ee
Here the second equality holds because logarithmic derivatives in $y$
and $s$ agree. The third equality relies on the fact that $\tilde{\cal P}_\zeta$
is a local quantity, as explained before. The averaging in the last term 
remains non-trivial because of the variation of $\bar{\zeta}$ over the large 
patch of size $L$ on which ${\cal P}_\zeta$ was originally defined. 

Expanding  to second order in $\bar{\zeta}$, we find
\be
{\cal P}_\zeta (y)\,=\,\tilde{\cal P}_\zeta (y)+\langle\bar{\zeta}\rangle
\frac{d}{d\ln y}\tilde{\cal P}_\zeta (y)
+\frac{1}{2}\,\langle\bar{\zeta}^2\rangle\,\frac{d^2}{d(\ln y)^2}
\tilde{\cal P}_\zeta (y)\,.\label{difps}
\ee
This quantifies the deviation of the observer-patch-dependent
spectrum ${\cal P}_\zeta$ from the locally defined spectrum
$\tilde{\cal P}_\zeta$ for large patches. It also clarifies the origin
of the large logarithms of $L$ found in the $\delta N$ formalism. They arise
because the expectation values of $\bar{\zeta}$ and $\bar{\zeta}^2$ grow
logarithmically with $L$. We note that, if the power spectrum 
${\cal \tilde P}_\zeta$ is scale independent, the two last terms in 
eq.~(\ref{difps}) vanish. This is compatible with eqs.~(\ref{eq:deltaN_final}),
(\ref{srpars}) and has a simple intuitive reason: The log-enhancement
occurs because in regions with a large background value of
$\bar{\zeta}$ the scale of a given mode is effectively misidentified. 
However, in the scale-invariant case, such a misidentification has no
consequences. 

We now compare eqs. (\ref{difps}) and (\ref{eq:deltaN_final})
quantitatively. To achieve this, we can rewrite derivatives in $\ln y$
in terms of derivatives in $\phi$ using $d\ln(y)/d\phi=N_\phi$ 
(alternatively we can directly use the results of Appendix B) 
and
express expectation values of $\bar{\zeta}$ and $\bar{\zeta}^2$ in
terms of $\phi$-correlators using eq.~(\ref{dnexp}). Equivalently, we can
observe that eq.~(\ref{difps}) is simply a second order Taylor 
expansion around the classical trajectory. Since there is an
unambiguous functional relation between fluctuations in $\phi$ and in 
$\zeta$ (the latter being equivalent to $\ln y$), we immediately
conclude that 
\be
{\cal P}_\zeta (y)={\cal \tilde  P}_\zeta (y)
+\frac12\,\frac{d^2\,{\cal \tilde  P}_\zeta}{d \phi^2}\,\langle 
\delta\phi^2\rangle={\cal \tilde P}_\zeta (y)
+\frac12\,{\cal P}_\chi\,\ln(L/y)\,
\frac{d^2  {\cal \tilde P}_\zeta 
}{d \phi^2}\,.
\ee
Here $\langle\delta\phi^2\rangle$ is the zero-momentum $\delta\phi$
correlator with UV cutoff $y$ and IR cutoff $L$, in analogy to $\langle
\bar{\zeta}^2\rangle$ above. Furthermore, we used the fact that the 
expectation value of $\delta \phi$ vanishes. Thus, we have achieved 
complete agreement with the IR divergences or, more precisely, with
the log-enhanced corrections which arise in the $\delta N$ formalism. 

To summarize, we have obtained a simple physical interpretation for
the logarithmic contributions to the power spectrum that were determined
in the previous section: The log-enhancement arises due to the use of
global coordinates in a very large region, where these deviate significantly 
from the invariant distance. It can be completely avoided if the power 
spectrum $\tilde{\cal P}_\zeta$ is considered, which is based on the
two-point correlation function defined in terms of the invariant
distance between each pair of points appearing in the spatial average.

As mentioned in the Introduction, a discussion closely related to the
present section appeared in \cite{Giddings:2010nc} while this paper
was being finalized. In particular, using an argument slightly
different from ours, a log-enhancement arising from the large-scale geometry 
of the reheating surface was derived. Our findings can be brought into 
agreement with \cite{Giddings:2010nc} if we drop the term 
$\sim\langle\bar{\zeta}\rangle$ in eq. (\ref{difps}) and set
$\alpha_\zeta=dn_\zeta/d \ln k$ to zero. This corresponds to keeping
just the $(n_\zeta-1)^2$ term in our eq.~(\ref{srpars}). Indeed, the
term proportional to $\langle\bar{\zeta}\rangle$ disappears if we 
rescale our coordinates on the reheating surface according to $y\to
ye^{-\bar{\zeta}}$. This can be directly checked from
eq. (\ref{difps}) and corresponds to the fact that, in these new
coordinates, the average of $\zeta$ vanishes. While this is
certainly a rather natural coordinate choice, it does not serve our
purpose of comparing with the $\delta N$ formalism: Indeed, in this
new coordinate system the value $\zeta=0$ does not any more correspond
to the endpoint of the classical trajectory. Nevertheless, setting 
questions related to the $\delta N$ formalism aside, the term $\sim\langle
\bar{\zeta}\rangle$ appears just to be matter of different conventions
between \cite{Giddings:2010nc} and our discussion. By contrast, we 
believe that our term $\sim\alpha_\zeta$ is real and, in general, no
more slow-roll suppressed than the $(n_\zeta-1)^2$ term. 

The agreement of the findings of this section (and of \cite{Giddings:2010nc},
subject to the caveats mentioned above) with the previously discussed 
calculation in the $\delta N$ formalism is a non-trivial result. In particular, 
\cite{Giddings:2010nc} argue that any attempt to understand the IR 
divergences using the $\delta N$ approach is incomplete due to certain 
divergences of the $\delta\phi$ correlation function which are already 
present at horizon crossing. One way of interpreting our modification of the 
$\delta N$ formalism is by saying that, via Eq.~(\ref{eq:scal_field_fluct}), 
we include such divergences in the formalism.\footnote{
We 
would like to thank Martin Sloth for
 a conversation helping us to 
 correct our comments on 
this issue, in the first version of the  paper.
}
It thus appears that our purely classical modification to the $\delta N$ 
formalism is sufficient to fully capture the leading logarithmic behaviour.

\section{Discussion}
We have studied IR divergences during inflation using both the $\delta N$
formalism and a simple, phenomenological approach based just on the geometry 
of the reheating surface. By implementing a simple modification of the 
$\delta N$ formalism, we took into account the effect of modes that left 
the horizon long before the scales we are observing on the Hubble scale. 
Including this effect provides new log-enhanced contributions to the power 
spectrum, at the same order in $H$ and slow-roll as the standard classical 
loop corrections. We found that the combination of all contributions can be 
assembled in an elegant formula, in which the log-enhanced contributions
are weighted by the second derivative of the tree level power spectrum,
with respect to the inflaton field.

This result can be understood intuitively by considering two power spectra: 
One is defined locally on the surface of reheating, using invariant 
distances to define the correlator. The other is based on the coordinate
distance on this surface and depends on global features of this surface,
in particular on long-wavelength modes. When expressed in terms of the
local spectrum, this latter, global spectrum exhibits an IR divergence 
associated to the size of the region on which it is measured. It is, 
in fact, this latter spectrum that is calculated in the $\delta N$ formalism
and the log-divergence found in both approaches is precisely the same.
This provides strong support for the modification of the $\delta N$ formalism 
we propose. In the case of an exactly scale invariant spectrum, the IR
logarithms are absent. For an observer dealing with a scale-dependent 
spectrum and having a very large region available for his measurement,
the use of the local spectrum, which is not affected by our IR effects,
appears to be clearly favored. 

In single-field, slow-roll inflation the coefficient multiplying the
logarithm is heavily suppressed. While it is usually also suppressed in 
multi-field models, this statement does not hold any more in complete 
generality. In fact, in some special cases the coefficient could be large 
enough to compensate for the power spectrum suppression found in 
single-field models \cite{Suyama:2008nt,Kumar:2009ge}. It would therefore 
clearly be of interest to extend our results to multi-field inflation. In 
these models, which contain isocurvature perturbations during inflation, the
Hubble scale will depend on several fields, even if the primordial
curvature perturbation depends on only one field. In cases where the 
loop correction is not small, it has been argued they could give an 
observable contribution to $\zeta$ through non-Gaussianity \cite{lr,non_gauss}, 
in particular through a special type of scale dependence of the bispectrum 
non-linearity parameter \cite{Suyama:2008nt,Kumar:2009ge}. For a discussion 
of the scale dependence of the tree level bispectrum see \cite{scale-dep}. 
It would thus be interesting to study the effect of our loop corrections on 
the bispectrum (and higher order $n$-point correlators).

\subsection*{Acknowledgements}
We would like to thank Nemanja Kaloper, David Seery and Christof
Wetterich for helpful discussions. This work was supported by the
German Research Foundation (DFG) within the Transregional Collaborative 
Research Centre TR33 ``The Dark Universe''. M.G. acknowledges support 
from the {\it Studienstiftung des Deutschen Volkes}. S.N.
acknowledges the funding from the Academy of Finland grant 130265.

\section*{Appendix A: Correlator of the curvature perturbation}
\label{ap:correlator_curv_perturbation}

In this appendix, we present the calculation 
relating the correlator of curvature perturbations $\zeta$ to 
correlators of the Gaussian random field $\chi$.
  The terms which need to be calculated are 
 \bea
  \label{eq:corr_curv_pert_1}
  \corrk{\zeta}{k}{p} & = \None^2 \corrk{\dphi}{k}{p}  +  \None \Ntwo \corrbase{ \dphi_{\vec{k}} \; \left(\dphi^2\right)_{\vec{p}} } \\
                & \phantom{=} + \frac{\Ntwo^2}{4} 
 \corrk{ \left(\dphi^2\right) }{k}{p} 
+ \frac{\None\Nthree}{3} \corrbase{ \dphi_{\vec{k}} \; \left(\dphi^3\right)_{\vec{p}} }\,.
\eea
In the previous equations, expressions like $(\delta \phi^2)_{\vec k}$
indicate convolutions. 
The terms in the last two lines contains correlators between
convolutions, that provide (see the first reference in 
\cite{non_gauss} for more
details)
\bea
  \label{eq:corr_curv_pert_2}
  \corrk{ \left(\dphi^2\right) }{k}{p} & =  4 \, \mathcal{P}_\chi \ln(kL) \,, \corrk{\chi}{k}{p} 
 \\
  \corrbase{ \dphi_{\vec{k}} \; \left(\dphi^3\right)_{\vec{p}} } & = 3 \, \mathcal{P}_\chi \ln(kL) \, \corrk{\chi}{k}{p} \quad .
\eea
Using the 
expansion of eq. (\ref{eq:scal_field_fluct}),
written in momentum space,  
the 2-point correlator of the scalar field $\delta \phi$ is given by
\bea
  \label{eq:corr_curv_pert_3}
  \corrk{\dphi}{k}{p} &= \corrk{\chi}{k}{p} \left[ 1 + \left( A_1^2 + 2 A_2 \right)
 \int_{l,m\,\ll\,k} \kmes{l} \kmes{m} \corrk{\chi}{l}{m} \right] \\
    &= \corrk{\chi}{k}{p} \left[ 1 + \left( A_1^2 + 2 A_2 \right) \mathcal{P}_\chi \ln(kL) \right]  \quad .
\eea
Note that, after applying Wick's theorem, only one term 
survives the conditions $l,m \ll k$. 
The remaining correlator $\corrbase{ \dphi_{\vec{k}} \; \left(\dphi^2\right)_{\vec{p}} }$,
appearing in the second term of eq. (\ref{eq:corr_curv_pert_1}),
 is  of order three in $\delta\phi$.
 Therefore, only terms at next-to-leading order, in the expansion of 
 eq.\ \eqref{eq:scal_field_fluct}, contribute to it. 
One finds
\bea
\langle \delta \phi_{\vec k}
\,( \delta \phi^2 )_{\vec{p}}
\rangle&=&A_1\,\int \frac{d^3 q}{(2\pi)^3} 
\int_{l\ll k} \frac{d^3 l}{(2\pi)^3}
\,\langle \chi_{\vec{k}-\vec{l}}\, \chi_{\vec l}\,\chi_{\vec{p}-\vec{q}}\,\chi_{\vec{q}}
\, \rangle
\\
&&+2 A_1\,\int \frac{d^3 q}{(2\pi)^3}
\int_{l\ll q} \frac{d^3 l}{(2\pi)^3}
\,\langle \chi_{\vec{k}}\, \chi_{\vec l}\,\chi_{\vec{q}-\vec{l}}\,\chi_{\vec{p}-\vec{q}}
\, \rangle.
\eea
Decomposing the four point functions appearing in the integrands
by means of  Wick's theorem, and taking care
of the conditions on the size of $l$, one finds 
\be
  \label{eq:corr_curv_pert_8}
 \corrbase{ \dphi_{\vec{k}} \; \left(\dphi^2\right)_{\vec{p}} } = 4 A_1 \mathcal{P}_\chi \ln(kL) \, \corrk{\chi}{k}{p} \,.
\ee
 Putting all terms together, the final 
result for the spectrum of curvature perturbations reads
\bea
  \label{eq:corr_curv_pert_9}
  \mathcal{P}_{\zeta}(k) & =  \mathcal{P}_\chi \left\lbrace \None^2 + \mathcal{P}_\chi \ln(kL) \left[ \left( A_1^2 + 2 A_2 \right) \None^2 + 4 A_1 \None\Ntwo + \Ntwo^2 + \None\Nthree \right] \right\rbrace \\
   & = \mathcal{P}_\chi \None^2 + \mathcal{P}_\chi \ln(kL) \; \frac{1}{2}  \frac{d^2}{d\phi^2}\left( \None^2 \mathcal{P}_\chi \right)\,.
\eea

\section*{Appendix B: Power spectrum  in coordinate
 space} \label{app:B}

In this appendix, we discuss properties of  the power spectrum 
in coordinate space, defined in  eq.~\eqref{psrs} as
\be
  \mathcal{P}_\zeta (y) \equiv - \frac{d}{d \ln{y}} \, \corrbase{ \zeta(\xv) \;
 \zeta(\xv+\yv) }  \quad .
\ee
The power spectrum in coordinate space is no more difficult to handle
than the usual power spectrum 
in momentum space. Moreover, as  discussed in the main text, 
it  allows for a simpler and  more direct analysis of log-enhanced
contributions associated with this quantity.

However, for the purpose of this appendix, we consider only observed
regions which are not much larger than $y$ (i.e. IR cutoffs not much
smaller than the relevant momentum scale $k$). Under this assumption, 
${\cal P}_\zeta$ and $\tilde{\cal P}_\zeta$ coincide and all that
follows applies to both spectra.

We can compare the two definitions of the power spectra as follows. 
 Fourier expanding the two-point function of curvature perturbations,
we can write
\bea\label{comp2ps}
 \mathcal{P}_\zeta (y)&=- \frac{d}{d \ln{y}}
\,\int \frac{d^3 p}{(2\pi)^3} \,\frac{d^3 k}{(2 \pi)^3}\,
e^{-i\left[\vec{p}\vec{x}+\vec{k}\left(\vec{x}+\vec{y} \right)\right]}
\,\langle\zeta(\vec{p}) \zeta(\vec{k})
\rangle\\
&=- \frac{d}{d \ln{y}}
\,\int \frac{k^2\,d k}{2}\,\int_{-1}^{1}d(\cos{\theta})\,
e^{i\,k\,y\,\cos{\theta}}\,{\cal P}_\zeta^{F} (k)
\\
&=- \frac{d}{d \ln{y}}
\,\int\limits_{ k_0} \frac{\sin(ky)}{ky} \; 
\mathcal{P}_\zeta^F(k) \; d\ln k \quad .
\eea
To pass from the first to the second line, we used the definition of 
the power spectrum in momentum space, given before 
eq.~(\ref{eq:deltaN_final}). It is labeled by an $F$, to distinguish 
it from the analogous quantity in real space. In the third line, 
we made the IR cutoff explicit introducing a scale $k_0$.
We can expand ${\cal P}_\zeta^F$ as
\be
{\cal P}_\zeta^F (k)\,=\,{\cal P}_\zeta^F (k_p)+
\ln{\left(\frac{k}{k_p}\right)}\,
\frac{d {\cal P}_\zeta^F}{d \ln{k}}\,(k_p)\,+\,\cdots\,,
\ee
where $k_p$ is some pivot scale, the inverse of which is of the order 
of $y$: we can set $y\simeq k_p^{-1}$. The second and higher terms are 
slow-roll suppressed with respect to the first term. Substituting this 
expansion in eq. (\ref{comp2ps}), we find 
\be\label{simppz}
 \mathcal{P}_\zeta (y)\,=\, {\cal P}_\zeta^F (k_p)\,
\frac{\sin{k_0 y}}{k_0 y}\quad+\quad
\mbox{slow-roll suppressed terms}
\ee
where the slow-roll suppressed terms are weighted by logarithms of order
 $\ln{(k_0 y)}$. 
We choose $k_0$ such that $k_0 y\simeq k_0/k_p\,\ll\,1$, but not as small
as to lead to large logarithms in the slow-roll suppressed terms. This implies
that the power spectra in coordinate and momentum space coincide, up to
negligible slow-roll corrections:
\be
{\cal P}_\zeta(y)\simeq {\cal P}_\zeta^F(1/y)\,.
\ee

\end{document}